\documentclass[12pt,here]{article}
\usepackage{epsfig}
\def\beqn{\begin{eqnarray}}
\def\eeqn{\end{eqnarray}}

\def\beq{\begin{equation}}
\def\eeq{\end{equation}}
\def\ba{\beq\new\begin{array}{c}}
\def\ea{\end{array}\eeq}

\def\lsim{\mathrel{\rlap{\lower3pt\hbox{\hskip0pt$\sim$}}
    \raise1pt\hbox{$<$}}}
\def\gsim{\mathrel{\rlap{\lower4pt\hbox{\hskip1pt$\sim$}}
    \raise1pt\hbox{$>$}}}

\def\Im{{\rm Im}}
\def\Re{{\rm Re}}

\newcommand {\slsh} [1] {\not{\hbox{\kern-2pt${#1}$}}}

\newcommand{\ntwo}{${\mathcal N}=2\;$}
\newcommand{\ntwot}{${\mathcal N}= \left(2,2\right)\; $}
\newcommand{\ntwoo}{${\mathcal N}= \left(0,2\right)\; $}
\newcommand{\none}{${\mathcal N}=1\;$}

\newcommand{\pt}{\partial}

\begin{document}


\begin{titlepage}

\begin{flushright}
FTPI-MINN-08/07, UMN-TH-2638/08\\
ITEP-TH-10/08\\
March 4, 2008
\end{flushright}

\begin{center}

{\Large \bf   Large-{\boldmath $N$} Solution of the Heterotic {\boldmath\ntwoo} \\ [1mm]
Two-Dimensional  {\boldmath$CP(N-1)$} Model }
\end{center}

\begin{center}
{\bf M.~Shifman$^{a}$ and \bf A.~Yung$^{a,b,c}$}
\end {center}
\vspace{0.3cm}
\begin{center}

$^a${\it  William I. Fine Theoretical Physics Institute,
University of Minnesota,
Minneapolis, MN 55455, USA}\\
$^{b}${\it Petersburg Nuclear Physics Institute, Gatchina, St. Petersburg
188300, Russia}\\
$^c${\it Institute of Theoretical and Experimental Physics, Moscow
117259, Russia}
\end{center}

\begin{abstract}

We continue explorations of non-Abelian strings,
focusing on the solution of a heterotic deformation of the $CP(N-1)$ model
with an extra right-handed fermion field and \ntwoo supersymmetry.
This model emerges as a low-energy theory on the worldsheet of
the BPS-saturated flux tubes (strings)
in \ntwo super\-symmetric QCD deformed by a
superpotential of a special type breaking
\ntwo supersymmetry down to \none$\!\!$.
Using  large-$N$ expansion we solve this model to the leading order in $1/N$.
Our solution exhibits spontaneous supersymmetry breaking for all values of the
deformation parameter. We identify the Goldstino field.
The discrete $Z_{2N}$ symmetry is shown to be
spontaneously broken down to $Z_2$; therefore  the worldsheet model has 
$N$ strictly degenerate vacua (with nonvanishing vacuum energy).
Thus, the heterotic $CP(N-1)$ model is in the deconfinement phase.
We can compare this dynamical pattern, on the one hand,
with the \ntwot $CP(N-1)$ model which has 
$N$ degenerate vacua with unbroken supersymmetry,  and, on the other hand, with 
nonsupersymmetric $CP(N-1)$ model with split quasivacua and the Coulomb/confining phase.
We determine the mass spectrum of the heterotic $CP(N-1)$ model in the large-$N$ limit.

\end{abstract}

\end{titlepage}

\newpage

\section{Introduction}
\label{intro}
\setcounter{equation}{0}

This paper continues exploration \cite{SY02} of the heterotic
\ntwoo model with the $CP(N-1)$ target space for the bosonic fields.
We solve this model at large $N$ in the leading order in $1/N$
using the $1/N$ expansion technique which was designed by Witten for a parent 
model \cite{W79}.

Two-dimensional  $CP(N-1)$ models emerge as effective low-energy theories on
the worldsheet of non-Abelian strings which had been found in a class of 
four-dimensional gauge theories \cite{HT1,ABEKY,SYmon,HT2}, see also the review
papers \cite{Trev,Etor,SYrev}. The main feature of the above non-Abelian strings
is the occurrence of orientational moduli associated with rotations of
their color fluxes inside a global SU($N$) group.  Internal dynamics of
the  orientational moduli is described by two-dimensional
$CP(N-1)$ model.

The first non-Abelian strings were found,
as critical solitonic solutions, in \ntwo supersymmetric gauge theories. 
These bulk theories have eight conserved supercharges; 
hence, four of them are conserved on the worldsheet. Thus, if the bulk
theory has \ntwo supersymmetry (SUSY),
the $CP(N-1)$ model on the string worldsheet is automatically \ntwo supersymmetric
(more exactly, it is \ntwot$\!\!$). 

Then it was shown that non-Abelian BPS-saturated strings\,\footnote{
We mean BPS-saturated at the classical level.} survive certain  \none preserving
deformations of the bulk \ntwo theory. In particular, a mass term for the
adjoint fields  was considered in \cite{SYnone}.
The string solution at the classical level remains BPS-saturated. With four supercharges in the bulk, normally, this would imply conservation of two supercharges on the string worldsheet. Previously it was believed, however, that
worldsheet supersymmetry gets an ``accidental" enhancement \cite{SYnone}.
This is due to the facts that ${\mathcal N} =(1,1)$ SUSY
is automatically elevated up to \ntwot on $CP(N-1)$ and, at the same time,
there are no ``heterotic" \ntwoo generalizations of the bosonic $CP(N-1)$ model. 

Edalati and Tong noted \cite{Edalati} that the target space is in fact
$CP(N-1)\times C$ rather than $CP(N-1)$.
If two fermionic moduli (former supertranslational moduli)
become coupled to superorientational moduli,
one {\em can} built a heterotic \ntwoo model with the $CP(N-1)$
target space for the bosonic moduli. Edalati and Tong suggested a 
general structure of such a model (in the gauged formulation).
Later Tong argued \cite{Tongd} that \ntwoo supersymmetry of the heterotic model
is spontaneously broken at the quantum level. 

A geometric representation for the heterotic
\ntwoo model was obtained in Ref.~\cite{SY02},
\newpage
\beqn
L_{{\rm heterotic}} && 
= 
\zeta_R^\dagger \, i\partial_L \, \zeta_R  + 
\left[\gamma\, g_0^2 \, \zeta_R  \, G_{i\bar j}\,  \big( i\,\partial_{L}\phi^{\dagger\,\bar j} \big)\psi_R^i
+{\rm H.c.}\right]
\nonumber
\\[4mm]
&&
 -g_0^4\, |\gamma |^2 \,\left(\zeta_R^\dagger\, \zeta_R
\right)\left(G_{i\bar j}\,  \psi_L^{\dagger\,\bar j}\psi_L^i\right)
\nonumber
\\[4mm]
&&
+G_{i\bar j} \big[\partial_\mu \phi^{\dagger\,\bar j}\, \partial_\mu\phi^{i}
+i\bar \psi^{\bar j} \gamma^{\mu} D_{\mu}\psi^{i}\big]
\nonumber
\\[4mm]
&&
- \frac{g_0^2}{2}\left( G_{i\bar j}\psi^{\dagger\, \bar j}_R\, \psi^{ i}_R\right)
\left( G_{k\bar m}\psi^{\dagger\, \bar m}_L\, \psi^{ k}_L\right)
\nonumber
\\[4mm]
&&
+\frac{g_0^2}{2}\left(1-2g^2_0|\gamma|^2\right)
\left( G_{i\bar j}\psi^{\dagger\, \bar j}_R\, \psi^{ i}_L\right)
\left( G_{k\bar m}\psi^{\dagger\, \bar m}_L\, \psi^{ k}_R\right)\,,
\label{cpn-1g}
\eeqn
where 
$G_{i\bar j}$ is the $CP(N-1)$ metric,
\beq
G_{i\bar j}=\frac{\partial^{2} K(\phi,\,\phi^{\dagger})}{\partial \phi^{i}\partial \phi^{\dagger\,\bar j}}\,,\quad K =\frac{2}{g_{0}^{2}}\ln\left(1+\sum_{i,\bar j=1}^{ N-1}\phi^{\dagger\,\bar j}\delta_{\bar j i}\phi^{i}\right)\,,
\eeq
$K$ is the K\"{a}ler potential, $\gamma$ is the deformation parameter, 
while the curvature tensor $R_{i\bar jk\bar l}$ 
can be written as
\beq
R_{i\bar{j} k\bar{m}} = - \frac{g_0^2}{2}\left(G_{i\bar{j}}G_{k\bar{m}} +
G_{i\bar{m}}G_{k\bar{j}}
\right)\,.
\eeq
The first two lines in Eq.~(\ref{cpn-1g}) describe the kinetic term and interactions
of an additional field, the right-handed fermion $\zeta_R$, which is the only remnant
of the $C$ factor of the moduli space of the string (i.e. $CP(N-1)\times C$).
In Ref.~\cite{SY02} $\gamma$  was obtained in terms of the deformation parameter
of the bulk theory. The last three lines describe the \ntwot $CP(N-1)$ model fields.
In fact, putting $\gamma$ to zero in the last line we get just
the standard \ntwot $CP(N-1)$ model.

Although qualitatively we agree with \cite{Edalati},
there are several  distinctions in a number of aspects.\footnote{These distinctions are
discussed in \cite{SY02}; we will say more on that below.}
Dynamics of the model (\ref{cpn-1g}) 
is intriguing and nontrivial; in particular, we {\em proved} \cite{SY02}
the fact that at small $|\gamma |$ supersymmetry is spontaneously broken, with $\zeta_R$ playing the role of Goldstino.
In the limit $|\gamma |\ll 1$
the vacuum energy density is proportional to the square of the
bifermion condensate,\footnote{The bifermion condensate in Eq.~(\ref{13})
must be evaluated at $\gamma =0$, i.e. in the \ntwot model.}
\beq
{\mathcal E}_{\rm vac} = |\gamma|^2\,
\left|\langle R_{i\bar j} \,  \psi_R^{\dagger\, \bar j} \,\psi_L^i\rangle
\right|^2 \neq 0\,.
\label{13}
\eeq
Thus, we confirmed Tong's conjecture  \cite{Tongd}  of spontaneous SUSY breaking.

Our task in the present paper is to solve the heterotic \ntwoo model
for arbitrary values of $\gamma$. Although we cannot do it for arbitrary $N,$
at large $N$ powerful methods of $1/N$ expansion
do allow us to find a complete solution. Qualitative features of this solution are expected to be valid even for $N=2$. The representation of the model which is most convenient
for the $1/N$ expansion is the so-called gauged formulation \cite{W79,W93}.

The model  (\ref{cpn-1g}) plays a two-fold role. If the deformation parameter is smaller than a critical value, to be discussed in Sect.~\ref{conformal}, it describes the worldsheet dynamics
of the four-dimensional heterotic string. If it becomes larger than a critical value, the string swells,
and two-derivative terms no longer capture its worldsheet dynamics.
In this limit the model (\ref{cpn-1g}) can be considered on its own right, with no reference
to non-Abelian strings in four dimensions. We focus on the first aspect.
At the same time,  dynamics of the heterotic $CP(N-1)$ model in the limit of infinitely
large deformation parameter (presumably, conformal) is intriguing and captivating.
This is a topic for a separate investigation, though.

Our main results are as follows.
We prove spontaneous SUSY breaking for all values of the deformation parameter,
identify the Goldstino field and find the mass spectrum of excitations.
The bifermion condensate is shown to develop at all finite values
of the deformation parameter.
It plays the role of the order parameter for the spontaneous breaking of the
$Z_{2N}$ symmetry of the heterotic $CP(N-1)$ model. $Z_{2N}$ is broken down to $Z_{2}$.
$N$ vacua of the model have nonvanishing energy but are strictly degenerate.
This fact guarantees that the model is in the deconfinement phase. 

Organization of the paper is as follows.
In Sect.~\ref{0,2} we briefly review the gauged formulation of the standard
\ntwot model, outline the gauged formulation of the
\ntwoo model and discuss various regimes for the deformation parameter.
In Sect.~\ref{oneloop} we calculate the one-loop effective potential in the large-$N$ limit,
and then analyze the vacuum structure. Section \ref{spectrum}
is devoted to the mass spectrum of excitations. In Sect.~\ref{compare}
we discuss deconfinement regime in the heterotic $CP(N-1)$ model,
as opposed to the Coulomb/confinement regime in its nonsupersymmetric ``parent."
Section \ref{conformal} presents arguments regarding the limiting dynamics of 
 the heterotic $CP(N-1)$ model in the limit of infinitely large deformation parameter.
 Section \ref{conclu} summarizes our findings. 

\section{Heterotic {\boldmath\ntwoo} model}
\label{0,2}
\setcounter{equation}{0}

We will start from reviewing the gauged formulation of the
conventional \ntwot mo\-del, and then elaborate a similar formulation for the
heterotic \ntwoo model. 

\subsection{Gauged formulation of the undeformed \ntwot model}

The target space of the orientational moduli of the non-Abelian string is
\cite{HT1,ABEKY,SYmon,HT2} 
\beq
 \frac{SU(N)_{C+F}}{SU(N-1)\times U(1)} \sim CP(N-1).
\eeq
With \ntwo SQCD in the bulk, orientational moduli 
completely decouple from the (super)translational ones, and can be
considered in isolation. Here we will briefly describe
the conventional  $CP(N-1)$ model with \ntwot supersymmetry 
(which is a part of string worldsheet theory) in the gauged formulation \cite{W93}.

This formulation implies introduction of 
a U(1) gauge field $A$ which gauges the U(1) symmetry
of $N$ complex fields $n^l$. The gauge coupling is 
assumed to be large in the bare Lagrangian, $e^2\to\infty$, so that the kinetic term of $A$ 
vanishes. The bosonic part of the action is
\beqn
S_{CP(N-1)\,{\rm bos}}
& =&
\int d^2 x \left\{
 |\nabla_{k} n^{l}|^2 +\frac1{4e^2}F^2_{kl} + \frac1{e^2}
|\pt_k\sigma|^2+\frac1{2e^2}D^2
\right.
\nonumber\\[3mm]
 &+&    2|\sigma|^2 |n^{l}|^2 + iD (|n^{l}|^2-r_0)
\Big\}\,,
\label{cpg}
\eeqn
where 
\beq
\nabla_k= \partial_k - i A_k 
\eeq
while $\sigma$ is a complex scalar
field. Moreover,  $r_0$ can be interpreted as a coupling constant, and $D$ is a $D$-component of the gauge multiplet. 
The $i$ factor in the last term of Eq.~(\ref{cpg})
is due to Euclidean notation. (For our conventions and notation see Ref.~\cite{SY02}
\footnote{The coupling constant $r_0$ is related to the coupling $\beta$
 used in \cite{SY02} as $r_0=2\beta$.}). 

The bare constant $r_0$ of the worldsheet model is related to the coupling 
constant of 
the bulk theory at the scale determined by the bulk gauge boson mass $ m_W$ 
(see e.g.~\cite{SYrev}),
\beq
r_0=\frac{4\pi}{g^2_2(m_W)} = \frac{N}{2\pi}\ln{\frac{m_W}{\Lambda}}\,,
\label{beta}
\eeq
where $\Lambda$ is the dynamical scale of  the $CP(N-1)$ model.
To keep the bulk theory weakly coupled we must assume
that
$
m_W\gg \Lambda \,.
$

Eliminating $D$ from the action (\ref{cpg}) leads to the constraint
\beq
|n^l|^2=r_0\,.
\label{nconstraint}
\eeq
As was mentioned, in the  limit $e^2\to\infty$
the gauge field $A_k$  and its \ntwo bosonic superpartner $\sigma$ become
auxiliary and can be eliminated by virtue of the equations of motion,
\beq
A_k =-\frac{i}{2r_0}\, \bar{n}_l \stackrel{\leftrightarrow}
{\partial_k} n^l \,.
\label{aandsigma}
\eeq
With $2N$ complex fields $n^l$, one real constraint (\ref{nconstraint}) and one phase
``eaten'' by gauging the common U(1) symmetry,
the model has $2N-1-1=2(N-1)$ real variables. This is precisely the number of the
bosonic fields in Eq.~(\ref{cpn-1g}).

Now, let us pass to the
fermionic sector of the \ntwot model.
The corresponding part of the action in the gauged formulation takes the form
\beqn
S_{CP(N)\,{\rm ferm}}
& =&
\int d^2 x \left\{
 \bar{\xi}_{lR}\,i\left(\nabla_{0}-i\nabla_3\right) \xi^{l}_R
+ \bar{\xi}_{lL}\,i(\nabla_{0}+i\nabla_3) \xi^{l}_L
 \right.
\nonumber\\[3mm]
 &+& 
\frac1{e^2}\,\bar{\lambda}_{R}\,i(\nabla_{0}-i\nabla_3) \lambda_R
+\frac1{e^2}\,\bar{\lambda}_{L}\,i(\nabla_{0}+i\nabla_3) \lambda_L
+ \Big [ i\sqrt{2}\,\sigma\,\bar{\xi}_{lR}\xi^l_L
\nonumber\\[3mm]
 &+& 
\left.
 i\sqrt{2}\,\bar{n}_l\,(\lambda_R\xi^l_L-
\lambda_L\xi^l_R) +\mbox{ H.c. } \Big]
\right\}\,,
\label{cpgf}
\eeqn
where the fields $\xi^l_{L,R}$ are the 
fermion superpartners of $n^l$ while $\lambda_{L,R}$
belong to the gauge multiplet. 
In what follows we will introduce a shorthand
\beq
\nabla_L \equiv \nabla_{0}-i\nabla_3\,,\qquad \nabla_R \equiv \nabla_{0} + i\nabla_3\,.
\eeq
Integrating the  fields $\lambda_{L,R}$ in the limit $e^2\to\infty$
we arrive at the following constraints:
\beq
\bar{n}^l \xi^l_L=0, \qquad \bar{n}^l \xi^l_R=0\, .
\label{nxiconstraint}
\eeq
Moreover, integrating over the $\sigma$ field  gives
\beq
\sigma=-\frac{i}{\sqrt{2}\,\, r_0}\;\bar{\xi}_{lL}\xi^l_R\,.
\label{sigma} 
\eeq
The U(1) gauge field $A$ now takes the form
\beqn
A_0+iA_3 
& = &
-\frac{i}{2r_0}\, \bar{n}_l \left(\stackrel{\leftrightarrow}{\partial_0}
+i\stackrel{\leftrightarrow}{\partial_3}\right) 
n^l -\frac{1}{r_0}\,\bar{\xi}_{lR}\xi_R^l\,,
\nonumber\\[3mm]
A_0-iA_3 
& = &
-\frac{i}{2r_0}\, \bar{n}_l \left(\stackrel{\leftrightarrow}{\partial_0}
-i\stackrel{\leftrightarrow}{\partial_3}\right) 
n^l -\frac{1}{r_0}\,\bar{\xi}_{lL}\xi_L^l\,.
\label{gaugefield} 
\eeqn
When we substitute the above expressions in the Lagrangian 
we generate the four-fermion interactions  
$\left( \bar{\xi}_{lL}\xi^l_R\right) \left( \bar{\xi}_{ k R}\xi^{ k}_L\right)$
and 
$\left( \bar{\xi}_{lL}\xi^{ l}_L\right) \left( \bar{\xi}_{ k R}\xi^k_R\right)$,
respectively.

Besides orientational and superorientational moduli, the
BPS-saturated non-Abelian strings have (super)translational moduli too. They are
related to the possibility of shifting the string center $x_{0i}$ in the plane orthogonal to
its axis, $i=1,2$.
The corresponding supertranslational moduli are $\zeta_R\,,\,\,\zeta_L$.
 In the \ntwo bulk theory
the worldsheet fields $x_{0i}(t,z)$, $\zeta_R(t,z)$  and $\zeta_L(t,z)$ are just
free fields  decoupled from the orientational sector. 

\subsection{Heterotic \ntwoo model}

After we break \ntwo supersymmetry of the bulk model by switching on the deformation
superpotential for the adjoint fields,
\beq
{\mathcal W}_{3+1}=({\mu}/{2})\, \left[{\mathcal A}^2
+  ({\mathcal A}^a)^2\right]
\label{defpo}
\eeq
(see \cite{SY02}), the above decoupling is no longer valid. 
The classical string solution still remains 1/2 BPS-saturated \cite{SYnone} (see also 
\cite{Edalati,SY02}). Two supercharges that survive on the string worldsheet still protect
$x_{0i}$ and $\zeta_L$. The worldsheet fields $x_{0i}(t,z)$ and $\zeta_L(t,z)$ remain
free fields decoupled from all others. This is no longer the case with regards to 
$\zeta_R$ which gets an interaction with $\xi$'s.

As a result, the heterotic \ntwoo model in the gauged formulation takes the form
\beqn
S
&=&
 \int d^2 x \left\{\,
\frac12 \, \bar{\zeta}_R \, i
\pt_L
\, \zeta_R
+  
\big[ \sqrt{2}i\,\omega\,\bar{\lambda}_{L}\,
\zeta_R  + {\rm H.c.}\Big]
\right.
\nonumber\\[3mm]
&+& 
 |\nabla_{k} n^{l}|^2 +\frac1{4e^2}F^2_{kl} + \frac1{e^2}
|\pt_k\sigma|^2+\frac1{2e^2}D^2
+    
2|\sigma|^2 |n^{l}|^2 + iD (|n^{l}|^2-r_0)
\nonumber\\[3mm]
 &+&
 \bar{\xi}_{lR}\,i \nabla_{L}\, \xi^{l}_R
+ \bar{\xi}_{lL}\,i\nabla_{R}\, \xi^{l}_L
+\frac1{e^2}\,\bar{\lambda}_{R}\,i \pt_{L}\, \lambda_R
+\frac1{e^2}\,\bar{\lambda}_{L}\,i \pt_{R}\, \lambda_L
\nonumber\\[3mm]
 &+& 
\left.
\Big[ i\sqrt{2}\,\sigma\,\bar{\xi}_{lR}\xi^l_L
+
 i\sqrt{2}\,\bar{n}_l\,(\lambda_R\xi^l_L-
\lambda_L\xi^l_R) + {\rm H.c.} \Big]
+
4\,|\omega|^2\,\left|\sigma
\right|^2  
\right\},\nonumber
\\
\label{cpg02}
\eeqn
where we omitted the fields  $x_{0i}(t,z)$ and $\zeta_L(t,z)$
as irrelevant for the present consideration. This is the action obtained in \cite{Edalati}.
Here $\pt_{L,R}=\pt_0\mp i\,\pt_3$.
The  terms containing $\zeta_R$ and/or $\omega$  break \ntwot  supersymmetry down 
to \ntwoo$\!\!$. The parameter $\omega$ is complex and 
dimensionless.\footnote{The relation of $\omega$ to the \ntwoo 
deformation parameter $\delta$ used in 
\cite{SY02} is $\omega=\sqrt{r_0}\;\delta$\,.}

Integrating over the axillary fields $\lambda$ we arrive at the constraints
\beq
\bar{n}^l \,\xi^l_L=0, \qquad \bar{\xi}_{lR}\, n^l=\omega
\,\zeta_R\,,
\label{modnxiconstraint}
\eeq
replacing those in Eq.~(\ref{nxiconstraint}).
We see that the constraint (\ref{nxiconstraint}) is modified for the right-handed
fermions $\xi_R$ implying that the supertranslational sector of the worldsheet theory is
no longer decoupled from the orientational one. The general structure of the deformation
in (\ref{cpg02}) is dictated by \ntwoo supersymmetry.

\subsection{On the value of the deformation parameter}

Edalati and Tong conjectured \cite{Edalati} that the worldsheet deformation
parameter $\omega$ is proportional
to the bulk deformation parameter $\mu$ (see Eq.~(\ref{defpo})),
\beq 
\omega\sim\mu\,.
\label{etconjecture}
\eeq
In the previous paper \cite{SY02} we derived the  worldsheet theory
(\ref{cpg02}) directly from the bulk theory. This derivation provides us with a relation
between the bulk and worldsheet deformation parameters, namely, 
\beq
\omega =
\left\{
\begin{array}{l}\rule{0mm}{5mm}
 {\rm const}\,\sqrt{r_0}\;\,\frac{g_2^2\mu}{m_W}\, , \quad\qquad\;  \mbox{small}\,\,\,
 \mu\,,\\[4mm]
 {\rm const}\,\sqrt{r_0\,\ln{\frac{g^2_2\mu}{m_W}}} \, , \qquad\, \mbox{large}\,\,\,\mu\,\, .
 \end{array}
 \right.
\label{omegasmu}
\eeq
Here $g_2^2$ is the SU(2) gauge coupling of the bulk theory.
The worldsheet deformation parameter $\omega$ is determined by the profile functions of the 
string solution \cite{SY02}. For simplicity we assume that $\mu$ and $\omega$ are real.
In the general case ${\rm arg \,\omega}={\rm arg \,\mu}$.

While the small-$\mu$  result in (\ref{omegasmu}) is in accordance with 
the Edalati--Tong conjecture, the large-$\mu$ behavior ($g_2^2\mu\gg m_W$
where $m_W$ is the $W$-boson mass)  indicated in 
Eq.~(\ref{omegasmu}) is in contradiction, since we get logarithmic rather than
power behavior.

The physical reason for the logarithmic behavior of the worldsheet deformation parameter
with $\mu$ is as follows. In the large-$\mu$ limit certain states in the bulk theory become
light \cite{SYnone,SY02}. This reflects the presence of the Higgs branch in \none SQCD
\cite{IS} to which our bulk theory flows in the  $\mu\to\infty$ limit. The argument of the 
logarithm in (\ref{omegasmu}) is the ratio of $m_W$ and a small  mass of the  light states
associated with this would-be Higgs branch \cite{SY02}.

Now, let us discuss $N$ counting. How all expressions relevant to the problem 
at hand depend on $N$ at 
large $N$? It is obvious that
\beq
r_0\sim N,
\label{rN}
\eeq
while the masses of physical states and the scale of the theory do not depend on $N$,
\beq
m_W\sim N^{0},\qquad g_2^2\mu \sim N^{0},\qquad \Lambda\sim N^{0}\,.
\label{massN}
\eeq
This, in turn, implies that the deformation parameter $\omega$ behaves 
as\,\footnote{In Ref. \cite{SY02} the bulk theory with the U(2) gauge group was studied.
Thus, strictly speaking, the result (\ref{omegasmu})  for the dependence
of $\omega$ on $\mu$ was derived only in the $N=2$ case. However, the $N$ dependence of $\omega$ is captured correctly by the factor $\sqrt{r_0}$ in these equations, therefore we  generalize here (\ref{omegasmu})  to arbitrary $N$.}
\beq
\omega\sim \sqrt{N}\,.
\label{omegaN}
\eeq

\subsection{Axial \boldmath{${\rm U(1)}$}}
\label{au1}

The model (\ref{cpg02}) has a U(1) axial symmetry which is broken by the chiral anomaly down 
to the discrete subgroup $Z_{2N}$ \cite{W79}.  Now,  the $\sigma$ field  is related to 
the fermion bilinear operator by the following formula: 
\beq
\sigma=-\frac{i}{\sqrt{2}(r_0+2|\omega|^2)}\;\bar{\xi}_{lL}\xi^l_R\,
\label{sigmamod} 
\eeq
(cf. (\ref{sigma})).
Moreover, under the above $Z_{2N}$ symmetry transformation it transforms  as 
\beq
\sigma \to e^{\frac{2\pi  k}{N}\,i}\, \sigma, \qquad k=1,..., N-1\,.
\label{ZN}
\eeq
We will show below that the $Z_{2N}$ symmetry is spontaneously broken by the condensation of $\sigma$, 
down to $Z_2$,
much in the same way as in the conventional \ntwot model \cite{W79}. 
This is equivalent to saying that
the fermion bilinear condensate $\langle \bar{\xi}_{lL}\xi^l_R\rangle$ develops, 
breaking the discrete $Z_{2N}$ symmetry down to $Z_2$. 

\section{ One-loop effective potential }
\label{oneloop}
\setcounter{equation}{0}

The \ntwot  model as well as nonsupersymmetric $CP(N-1)$ model
were solved by Witten in the large-$N$ limit \cite{W79}.
The same method was used in
\cite{GSYphtr} to study nonsupersymmetric $CP(N-1)$ model with twisted mass. In this section we will generalize Witten's analysis 
to solve the \ntwoo theory. 

Since the action (\ref{cpg02}) is quadratic in the fields $n^{l}$ and $\xi^l$
we can integrate over these fields and then minimize the resulting
effective action with respect to the  fields from the gauge multiplet. The large-$N$ limit ensures the corrections to the saddle point approximation to be  small. In fact,
this procedure boils down to calculating a small set of one-loop graphs with the
$n^{l}$ and $\xi^l$  fields propagating in loops.
After integrating $n^{l}$ and $\xi^l$ out, we must check self-consistency.

Integration over $n^{l}$ and $\xi^l$ in (\ref{cpg02})
yields the following determinants:
\beq
\left[ {\rm det}\, \left(-\pt_{k}^2 +iD
+2|\sigma|^2\right)
\right]^{-N}
\left[ {\rm det}\, \left(-\pt_{k}^2 
+2|\sigma|^2\right)
\right]^{N},
\label{det}
\eeq
where we dropped the gauge field $A_k$. The first determinant here comes from the
boson loops while the second from fermion loops. Note, that the $n^{l}$ mass 
 is given by $iD+2|\sigma|^2$ while that of fermions $\xi^l$
is $2|\sigma|^2$. If supersymmetry is unbroken (i.e.  $D=0$) these masses are equal,
and the product of the determinants reduces to unity, as
it should be.

Calculation of the determinants in Eq.~(\ref{det}) 
is straightforward. We easily get the following contribution to the
effective action:
\beqn
\frac{N}{4\pi}\left\{\left(iD
+2|\sigma|^2\right)
\left[\ln\, {\frac{M_{\rm uv}^2}{iD
+2|\sigma|^2}}+1\right]
-2|\sigma|^2
\left[\ln\, {\frac{M_{\rm uv}^2}{
2|\sigma|^2}}+1\right]
\right\},
\label{detr}
\eeqn
where quadratically divergent contributions from bosons and fermions do
not depend on
$D$ and $\sigma$ and cancel each other. Here $M_{\rm uv}$ is an ultraviolet (UV) cutoff.
 Remembering that 
the action in (\ref{cpg02}) presents an effective low-energy theory on
the string worldsheet one can readily identify the  UV cutoff  in terms of  bulk parameters,
\beq
M_{\rm uv}=m_W \,.
\label{Muv}
\eeq
Invoking Eq.~(\ref{beta}) we conclude that the bare coupling constant
 $r_0$ in (\ref{cpg02}) can be parameterized as
\beq
r_0=\frac{N}{4\pi}\, \ln\, {\frac{M_{\rm uv}^2}{\Lambda^2}}\,.
\eeq
Substituting this expression in (\ref{cpg02}) and adding 
the one-loop correction
(\ref{detr})
we see that the term proportional to 
$iD \ln\, {M_{\rm uv}^2}$ is canceled out, and the effective action is
expressed in terms of the renormalized coupling constant,
\beq
\rule{0mm}{8mm}
r_{\rm ren}=\frac{N}{4\pi}\, 
\ln\, {\frac{iD +2|\sigma|^2}{\Lambda^2}}\, .
\label{coupling}
\eeq

\vspace{2mm}

Assembling  all contributions together we get the effective potential as 
a function of the $D$ and $\sigma$ fields  in
the form
\beqn
 V_{\rm eff} &=& \int d^2 x \,\,\frac{N}{4\pi}\,
\left\{
 -\left(iD+2|\sigma|^2\right)
\ln\, {\frac{iD+2|\sigma|^2}{\Lambda^2}} +iD\right.
\nonumber\\[3mm]
&+& \left.
2|\sigma|^2\,\ln\, {\frac{2|\sigma|^2}{\Lambda^2}}
+2|\sigma|^2\,u\, \right\}\,,
\label{effpot}
\eeqn
where instead of the deformation parameter $\omega$ we introduced a more 
convenient (dimensionless) parameter $u$ which does not scale with
$N$,
\beq
u=\frac{8\pi}{N}\,|\omega|^2,
\label{u}
\eeq
see Eq.~(\ref{omegaN}).

Minimizing this potential with respect to $D$ and
$\sigma$ we arrive at the following relations:
\beqn
&&
r_{\rm ren}=\frac{N}{4\pi}\, 
\ln\, {\frac{iD +2|\sigma|^2}{\Lambda^2}}=0\,,
\nonumber \\[5mm]
&&
\ln\, {\frac{iD +2|\sigma|^2}{2|\sigma|^2}}=u\,.
\label{veq}
\eeqn
Equations (\ref{veq}) represent our {\em master set} which
determines the vacua of the theory.
Solutions can be readily found,
\beqn
&&
2|\sigma|^2=\Lambda^2\,e^{-u}\,,\qquad \sigma = \frac{1}{\sqrt 2}\,\Lambda\, 
\exp\left( -\frac{u}{2} + \frac{2\pi\,i\, k}{N}
\right), \quad k=0, ..., N-1,
\nonumber \\[3mm]
&&
iD = \Lambda^2\,\left(1-e^{-u}\right)\,.
\label{sol}
\eeqn
The phase factor of $\sigma$ does not follow from (\ref{veq}),
but we know of its existence from the fact of the spontaneous
breaking of the discrete chiral $Z_{2N}$ down to $Z_2$, see Sect.~\ref{au1}.
Substituting this solution in Eq.~(\ref{effpot}) we get the expression for the vacuum energy
density,
\beq
{\mathcal E}_{\rm vac}= \frac{N}{4\pi}\,iD=\frac{N}{4\pi}\,\Lambda^2\,\left(1-e^{-u}\right)\,.
\label{vacenergy}
\eeq
\vspace{1mm}

\noindent
Note that at small $u$ the  vacuum energy
density reduces to
\beq
{\mathcal E}_{\rm vac} \propto u\,N\, |\sigma |^2\, ,
\label{vae}
\eeq
in full accord with Eq.~(\ref{13}). On the other hand, at large $u$
\beq
{\mathcal E}_{\rm vac}\to \frac{N}{4\pi}\,\Lambda^2\,.
\eeq
This value is of the order of $N\Lambda^2$. Needless to say, the linear $N$
dependence was expected.

It is instructive to discuss the first condition in (\ref{veq}). 
That $r_{\rm ren}=0$ was a result of Witten's analysis \cite{W79} too. 
This fact, $r_{\rm ren}=0$,
implies that
in quantum theory (unlike the classical one)
\beq
\langle\,  |n^l|^2\, \rangle =0\, ,
\label{nohiggs}
\eeq
i.e. the global SU$(N)$ symmetry is not spontaneously broken 
in the vacuum and, hence,  there are no massless Goldstone bosons.
All bosons get a mass.

If the deformation parameter $u$ vanishes, the vacuum energy vanishes too and 
supersymmetry is not broken, in full
accord with Witten analysis \cite{W79} and with the fact that
the Witten index is $N$ in this case \cite{WI}. The $\sigma$ field develops
a vacuum expectation value (VEV) breaking $Z_{2N}$ symmetry 
(\ref{ZN}).\footnote{The vacuum structure (\ref{sol}) of the \ntwot model at $u=0$ was also
obtained by Witten for arbitrary $N$ in \cite{W93} using a superpotential of 
the Veneziano--Yankielowicz type \cite{Veneziano}.} 
As we switch on the deformation 
parameter $u$, the $D$ component develops a VEV; hence, \ntwoo supersymmetry is
spontaneously broken. The vacuum energy density no longer vanishes. 

In the limit $\mu\to \infty$, the deformation
parameter $u$ behaves logarithmically with $\mu$,
\beq
u\,= {\rm const}\left(\ln\, {\frac{m_W}{\Lambda}}\right)  \left(\ln\, {\frac{g^2_2\mu}{m_W}}\right),
\label{umularge}
\eeq
where the constant above does not depend on $N$.
At any finite $u$ the 
$\sigma$-field condensate does not vanish, labeling  $N$ distinct vacua as
indicated in Eq.~(\ref{sol}). In each vacuum
$Z_{2N}$ symmetry is spontaneously broken  down to $Z_2$;  
 the order parameter is $\langle\,\sigma\,\rangle$.
We will discuss physics of the model in the large-$\mu$ limit in more detail
in Sects.~\ref{compare} and \ref{conformal}.

\begin{figure}
\epsfxsize=7cm
\centerline{\epsfbox{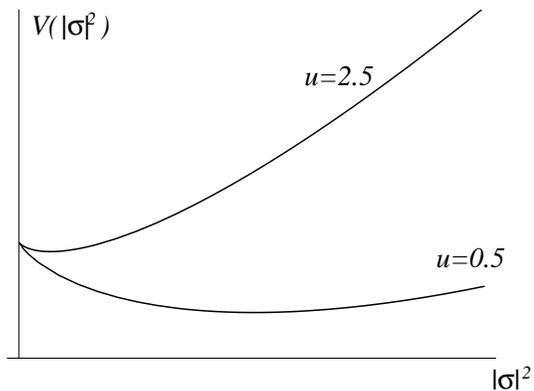}}
\caption{\small The potential (\ref{sigmapot}) as a function of $|\sigma  |^2$. }
\label{xxx}
\end{figure}

To conclude this section let us integrate over the axillary field $iD$ in 
(\ref{effpot}) to get the effective potential of the \ntwoo model as a function of the physical field $\sigma$,
\beq
V(\sigma)=\frac{N}{4\pi}\left\{
\Lambda^2 + 2|\sigma|^2\left[u+\ln\, {\frac{2|\sigma|^2}{\Lambda^2}}-1\right]\right\}.
\label{sigmapot}
\eeq
Clearly, the minimum of this potential is at $|\sigma |^2$ given in the first line
of Eq.~(\ref{sol}). The plot in Fig.~\ref{xxx}
illustrates the tendency of the growth of ${\mathcal E}_{\rm vac}$ and decrease of
$\langle\,\sigma\,\rangle$ as the deformation parameter $u$ increases.

\section{The physical spectrum}
\label{spectrum}
\setcounter{equation}{0}

Our next task is calculation of the mass spectrum of the theory (\ref{cpg02}).
To this end we start from the
one-loop effective action and analyze it as a function of  both the ``extra" 
field $\zeta_R$ and 
the boson and fermion fields from the gauge supermultiplet
($A_k$, $\sigma$ and $\lambda$). 
After integration over $n^{l}$ and $\xi^l$ we obtain
\newpage

\beqn
S_{\rm eff}
&=&
 \int d^2 x \left\{
\frac1{4e_{\gamma}^2}F^2_{kl} + \frac1{e_{\sigma 1}^2}
|\pt_k({\rm Re}\,\sigma )|^2 +\frac1{e_{\sigma 2}^2}
|\pt_k({\rm Im}\,\sigma )|^2
\right.
\nonumber\\[3mm]
 &+&    
\frac1{e^2_{\lambda}}\,\bar{\lambda}_{R}\,i\,\pt_{L}\,  \lambda_R
+\frac1{e^2_{\lambda}}\,\bar{\lambda}_{L}\,i\, \pt_{R}\, \lambda_L
+\frac12 \, \bar{\zeta}_R \, i
\, \pt_L \, \zeta_R
\nonumber\\[3mm]
 &+& 
\left.
V(\sigma)+i\frac{N}{\pi}\,\frac{\Im\,\sigma}{|\sigma|}\,F^{*}
 +\left[ i\sqrt{2}\,\Gamma\,\bar{\sigma}\,\bar{\lambda}_{L}\lambda_R  
+ \sqrt{2}\,i\,\omega\,\bar{\lambda}_{L}\,
\zeta_R  +{\rm H.c.}\right] \right\},
\label{41}
\eeqn
where 
it is anticipated that in the   vacuum under consideration 
we will have Im$\,\sigma =0$, i.e. we consider the vacuum given by Eq.~(3.9) with $k=0$. This condition determines the form of the 
$F^*\sigma$ coupling, namely Im$\sigma \, F^*$. If necessary, it is not difficult to modify the expression (\ref{41}) for other vacua.
Moreover,
\beq
\pt_L\equiv \pt_{0}-i\pt_3\,,\qquad \pt_R\equiv \pt_{0}+ i\pt_3\,,
\eeq
$V(\sigma)$ is given in Eq.~(\ref{sigmapot}) while $F^{*}$ is the  dual gauge field strength,
\beq
F^{*}=\frac12\varepsilon_{kl}F_{kl}\,.
\eeq
Here $e^2_{\gamma}$, $e^2_{\sigma}$ and $e^2_{\lambda}$ are the coupling constants which
determine the wave function renormalization for  the photon, $\sigma$,
and $\lambda$ fields,
respectively. Moreover, $\Gamma$ is the induced Yukawa coupling. 
These couplings are given by one-loop graphs which we will consider below.

The $({\rm Im}\,\sigma)\, F^{*}$ mixing was calculated by Witten in \cite{W79}
for \ntwot theory. This mixing  is due to
the chiral anomaly which makes the photon massive in two dimensions. In the effective action
this term is represented by the mixing of the gauge field with the imaginary part of 
$\sigma$. Since the anomaly is not modified by \ntwot breaking deformation, we can use 
Witten's result in the deformed theory.

The wave function renormalizations of  the fields from the gauge supermultiplet are, 
in principle, momentum-dependent. We calculate them below in the low-energy limit assuming 
the external momenta to be small.

\begin{figure}
\epsfxsize=6cm
\centerline{\epsfbox{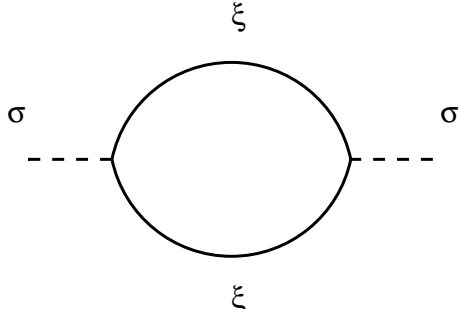}}
\caption{\small
The wave function renormalization for  $\sigma$. }
\label{fig:esigma}
\end{figure}

The wave function renormalization for $\sigma$ is given by the $n$ loop graph
and a similar graph with the $\xi$ fermions (Fig. 2). A straightforward calculation yields
\beqn
\frac1{e^2_{\sigma 1}}
&=&
\frac{N}{4\pi}\,
\frac{1}{2|\sigma |^2}\left(\frac13 + \frac23\,\frac{|\sqrt{2}\sigma |^4}{(2|\sigma |^2 +iD)^2}
\right) ,
\nonumber\\[3mm]
\frac1{e^2_{\sigma 2}}
&=&
\frac{N}{4\pi}\,
\frac{1}{2|\sigma |^2}\,.
\label{esigma}
\eeqn
The integral is saturated at momenta of the order of $\xi$
mass $\sqrt{2}|\sigma|$.

\begin{figure}
\epsfxsize=10cm
\centerline{\epsfbox{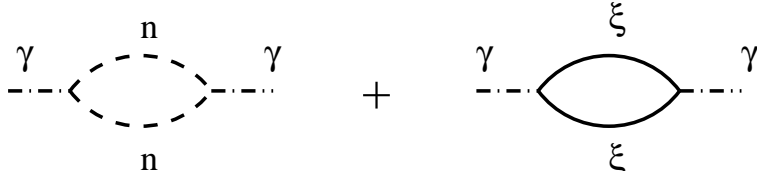}}
\caption{\small
The wave function renormalization for the gauge field.}
\label{fig:photon}
\end{figure}

The wave function renormalization for the gauge field was calculated by Witten in 
\cite{W79}. The result is 
\beq
\frac1{e^2_{\gamma}}=\frac{N}{4\pi}\,\left[\frac13\,\frac{1}{iD+2|\sigma|^2}+
\frac23\,\frac{1}{2|\sigma|^2}\right].
\label{egamma}
\eeq
The right-hand side in Eq.~(\ref{egamma}) is given by two graphs in 
Fig.~\ref{fig:photon}, with bosons $n^l$ and fermions $\xi^l$
in the loops. The first term in (\ref{egamma}) comes from bosons while the second
one is due to fermions.

\begin{figure}
\epsfxsize=5cm
\centerline{\epsfbox{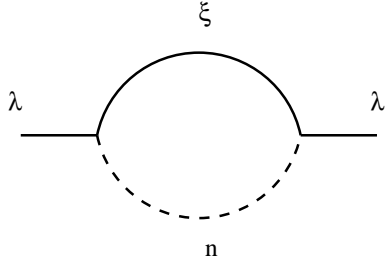}}
\caption{\small
The wave function renormalization for  $\lambda$.}
\label{fig:lambda}
\end{figure}

The renormalization for the $\lambda$
fermions is shown in Fig.~\ref{fig:lambda}. This graph
gives
\beq
\frac1{e^2_{\lambda}}=2\,\frac{N}{4\pi}\,\int d k^2
\frac{2|\sigma|^2}{(k^2+iD+2|\sigma|^2)(k^2+2|\sigma|^2)^2}\,.
\label{elambda}
\eeq
Note that the gauge supermultiplet was introduced in (\ref{cpg02}) as  axillary fields, with no kinetic
terms ($e^2\to\infty$ in (\ref{cpg02})). We see that the kinetic terms for these fields are generated at the one-loop level. Therefore, these fields become physical \cite{W79}.

\begin{figure}
\epsfxsize=5cm
\centerline{\epsfbox{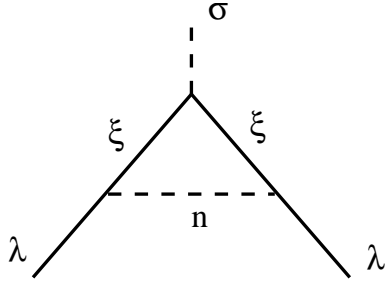}}
\caption{\small
The induced Yukawa vertex.}
\label{fig:Gamma}
\end{figure}

The Yukawa coupling is determines 
by the one-loop graph in Fig.~\ref{fig:Gamma} which gives
\beq
\Gamma=2\,\frac{N}{4\pi}\,\int d k^2
\frac{1}{(k^2+iD+2|\sigma|^2)(k^2+2|\sigma|^2)}\,,
\label{Gamma}
\eeq
where one propagator comes from the bosons $n^l$ while the other from 
the fermions $\xi^l$.

The subsequent analysis is straightforward if we limit ourselves to two limits:
small and large values of the bulk deformation parameter $\mu$.
These are the limits which were considered in \cite{SY02}.

\subsection{Small-{\boldmath $\mu$} limit}
\label{smamu}

Let us first put $\omega=0$ and reproduce the mass spectrum of the theory in the \ntwot
limit that had been obtained by Witten in \cite{W79}. If  $\omega=0$ then $D=0$ and  supersymmetry is 
unbroken. Masses of the $n^l$ bosons  and $\xi^l$ fermions coincide. They are given by 
the following formula:
\beq
m_{n}=m_{\xi}=\sqrt{2}|\sigma|=\Lambda\,,
\label{nmass22}
\eeq
where we used Eq.~(\ref{sol}) at $u=0$.
The wave function renormalizations are also equal in this limit,
\beq
\frac1{e^2_{\sigma}}=\frac1{e^2_{\gamma}}=\frac1{e^2_{\lambda}}=
\frac{N}{4\pi}\,\frac{1}{2|\sigma|^2}=\frac{N}{4\pi}\,\frac{1}{\Lambda^2}\,,
\label{e22}
\eeq
while the Yukawa coupling (\ref{Gamma}) is
\beq
\Gamma=\frac{N}{4\pi}\,\frac{2}{\Lambda^2}\,.
\label{Gamma22}
\eeq
Since $\sigma$ develops a VEV, the Yukawa term in (\ref{41}) gives a mass to the
$\lambda$ fermion. Using (\ref{e22}) and (\ref{Gamma22}) we get
\beq
m_{\lambda_{R}}=m_{\lambda_{L}}=2\sqrt{2}|\sigma|=2\Lambda\,.
\label{lmass22}
\eeq
Minimizing the potential $V(\sigma)$ in Eq.~(\ref{sigmapot}) we calculate the mass of the real
part of the $\sigma$ field,
\beq
m_{\Re\,\sigma}=2\Lambda\,,
\label{smass22}
\eeq
where we also used Eq.~(\ref{e22}).
The anomalous $\left(\Im\,\sigma \right) F^{*}$ mixing in (\ref{41}) gives masses to both photon
and the imaginary part of $\sigma$. Using (\ref{e22}) we get
\beq
m_{\rm ph}=m_{\Im\,\sigma}=2\Lambda\,.
\label{phmass22}
\eeq
\vspace{1mm}

We see that all fields from the  gauge multiplets have the same mass $2\Lambda$ in accordance  with \ntwot supersymmetry.
The factor of 2 in Eq.~(\ref{phmass22}) is easy to understand if we take into account that, say, $\lambda$
is a bound state of $n$ and $\xi$, each of them has mass $\Lambda$
and all interactions are $O(1/N)$. The binding energy
$O(1/N)$ is not seen in the leading order in the large-$N$ expansion.
The field $\zeta_R$ is massless and sterile at $\omega=0$.

Now let us switch on a small deformation parameter $u$. 
The field $\zeta_R$ is no longer sterile. This  explicitly breaks \ntwot supersymmetry down to 
\ntwoo$\!\!$. Moreover,  \ntwoo supersymmetry gets spontaneously broken due to 
the VEV
of  the $D$ component in (\ref{sol}). The vacuum energy no longer vanishes, it becomes
proportional to the deformation parameter $u$, see Eq.~(\ref{vae}). The spectrum
of fields from the former \ntwot gauge multiplet does not change very much:
superpartners split acquiring mass differences
linear in $u$ around the average value $2\Lambda$.

Due to the spontaneous supersymmetry breaking we have a massless Goldstino fermion in the theory.
To check this explicitly we diagonalize the mass matrix for the $\zeta_R$, $\lambda_R$ and 
$\lambda_L$ fermions in Eq.~(\ref{41}). Equating the determinant of this matrix to zero
produces the following equation for the mass eigenvalues $m$:
\beq
m^3-m\left[ 2|\sigma|^2 \,\Gamma^2\,e_{\lambda}^4+4\,\omega^2\,e_{\lambda}^2\right]=0\,.
\label{detmass}
\eeq
At any $\omega$ we have a {\em vanishing} eigenvalue. It corresponds to a massless 
Goldstino. Clearly, at small $\omega$ this fermion coincides 
with $\zeta_R$
(with an $O(\omega)$ admixture of the $\lambda$ fermions). At small $\omega$ we can neglect 
the second term in the square brackets
in Eq.~(\ref{detmass}). Then substituting (\ref{e22}) and (\ref{Gamma22}) into the first term
we reproduce the result (\ref{lmass22}) for the masses of the $\lambda$ fermions.

\subsection{Large-{\boldmath $\mu$} limit}

As we increase the bulk deformation parameter $\mu$ so does the worldsheet deformation
parameter $u$.
Spontaneous supersymmetry breaking in the worldsheet model gets
stronger (the strings are no 
longer BPS). The \ntwoo super\-multiplet splittings grow.

In this regime the masses of the $n^l$ bosons and $\xi^l$ fermions become essentially 
different. They are
\beq
m_{n}= \sqrt{ iD+{2}|\sigma|^2 }=\Lambda,\quad m_{\xi}=\sqrt{2}|\sigma|=
\Lambda\,\exp{\left(-\frac{u}{2}\right)},
\label{nmass02}
\eeq
where we used Eqs.~(\ref{sol}). The fermions are much lighter than their 
bosonic counterparts. The mass split is unsuppressed by $1/N$ since it is due to
$\xi\,\zeta\, n$ coupling which is of order unity in the
regime under consideration.

The mass of the real part of $\sigma$ can be readily calculated using the potential $V(\sigma)$ 
(see (\ref{sigmapot})),  
\beq
m_{\Re\,\sigma}=2\Lambda\,\exp{\left(-\frac{u}{2}\right)}\,\left\{
\frac13 + \frac23\,\, e^{-2u}
\right\}^{-1/2}
\,,
\label{smass02}
\eeq
where we also invoked Eq.~(\ref{esigma}).
Moreover, diagonalizing the  photon-$\Im\,\sigma$ mixing  in Eq.~(\ref{41}) 
we get 
\beq
m_{\rm ph}=m_{\Im\,\sigma}=\sqrt{6}\,\Lambda\,\exp{\left(-\frac{u}{2}\right)}
\,.
\label{phmass02}
\eeq
The binding between the constituents is again due to
$\xi\,\zeta\, n$ coupling and is unsuppressed by $1/N$.

The above masses no longer coincide with the mass of the real part of
$\sigma$. Technically the difference arises due to the difference in the
coupling constants
$e_{\sigma}$ in (\ref{esigma}) and $e_{\gamma}$ in (\ref{egamma}). When 
the $\sigma$ VEV  is
small the second term in (\ref{egamma}) dominates, and $e_{\gamma}$ becomes
\beq
\frac1{e^2_{\gamma}}=\frac{N}{6\,\pi} \,\frac{1}{2|\sigma|^2}=\frac{N}{6\pi}\,\frac{1}{\Lambda^2}
\,e^{u}\,\,,
\label{egamma02}
\eeq
while $e_{\sigma}$ is  given by Eq.~(\ref{esigma}).

The fermion masses can be obtained from Eq.~(\ref{detmass}). Clearly, at large $\mu$ the 
$\lambda_L\, \zeta_R$ mixing dominates in Eq.~(\ref{41}). In this limit
$\lambda_R$ (the bound state of $\xi$ and $n$) becomes the
massless Goldstino state,
\beq
m_{\lambda_{R}}=0\,,
\label{goldstino}
\eeq
while the masses of $\lambda_L$ and $\zeta_R$ are given by non-zero roots 
of Eq.~(\ref{detmass}),
\beq
m_{\lambda_{L}}=m_{\zeta_{R}}=\Lambda\,\sqrt{u}\, \, ,
\label{lmass02}
\eeq
where we used the fact that the coupling $e_{\lambda}$ in (\ref{elambda}) reduces in this limit to
\beq
\frac1{e^2_{\lambda}}=\frac{N}{4\pi}\,\frac{2}{\Lambda^2}\, .
\label{elambda02}
\eeq

We see that these two fermions become heavy in the limit $u\gg 1$. Thus,  the low-energy
effective theory contains the  light (but massive!) photon, two light $\sigma$ states and 
only fermion: the massless 
Goldstino $\lambda_R$.

\section{ Kink deconfinement vs. confinement}
\label{compare}
\setcounter{equation}{0}

As was already mentioned, both the \ntwot and nonsupersymmetric $CP(N-1)$ models were 
solved by Witten at large $N$ \cite{W79}. 
Witten showed that $n$'s are in fact kinks, $\xi$'s their superpartners,
and they are confined in nonsupersymmetric
version while adding supersymmetry
converts confinement into deconfinement.
This is in one-to-one correspondence with the existence of
$N$ {\em degenerate} vacua in the latter case. These vacua become {\em non}degenerate quasivacua
in nonsupersymmetric $CP(N-1)$ models \cite{GSYphtr}.

In this section we will compare the large-$N$ solutions
for all three theories: with \ntwot and  \ntwoo supersymmetry as well as
 the nonsupersymmetric version.

One common feature of all three cases is that spontaneous breaking of the global SU$(N)$
(flavor) symmetry present  at the classical level disappears when quantum
effects are taken into account. There are no massless Goldstone bosons in the physical
spectra of all three theories. The $n^l$ fields acquire mass of the order of $\Lambda$.
Another common feature is that the U(1) gauge field introduced as an axillary field at the 
classical level develops a kinetic energy term and becomes propagating.

In the \ntwot $CP(N-1)$ model supersymmetry is not spontaneously broken and 
the model has $N$ strictly degenerate vacua. The order parameter which characterizes these
vacua is the vacuum expectation value of the $\sigma$ field  given in the
first line of Eq.~(\ref{sol}) for $u=0$,  or, which is the same, the bifermion 
condensate (\ref{sigma}).

Since we have $N$ different vacua and $Z_{2N}$ symmetry is spontaneously broken 
down to $Z_2$ we have kinks
interpolating between these vacua. These kinks are described by the fields $n^l$ belonging to
the fundamental representation of the SU$(N)$ group \cite{W79,HoVa}. 
From the standpoint of the underlying bulk
theory these kinks are interpreted as confined monopoles \cite{Tong,SYmon,HT2}. In the bulk 
theory we have the Higgs phase; thus, the monopoles 
are confined in the four-dimensional sense,
i.e. they are attached to strings.\footnote{Confinement in two dimensions is
confinement along the string.}
 It is easy to show that the values of the magnetic 
charges of the monopoles from the SU$(N)$ subgroup of the gauge group ensure that these monopoles are the 
string junctions of two elementary non-Abelian strings, see the review 
paper \cite{SYrev} for details.

As was shown above, in the \ntwoo theory  supersymmetry is spontaneously 
broken. The vacuum energy density does not vanish, see (\ref{vacenergy}). This means that strings
under consideration
are no longer BPS and their tensions get a shift (\ref{vacenergy}) with respect to the classical
value $T_{\rm cl}=2\pi\xi$. However, this shift is the {\rm  same} for all $N$ elementary strings.
Their tensions are strictly degenerate; $Z_{2N}$ symmetry is spontaneously broken
down to $Z_2$. The 
order parameter (the $\sigma$ field VEV) remains nonvanishing at any finite value of 
 the bulk parameter $\mu$.

The kinks that interpolate between different vacua of the worldsheet theory are described by 
the  $n^l$ fields. Their masses are given in Eq.~(\ref{nmass02}). In \ntwoo theory
the  masses of the boson and
fermion superpartners are split. The bosonic kinks have masses
$\sim \Lambda$ in the large-$\mu$ limit,
while the fermionic kinks become light. Still their masses remain finite 
and nonvanishing at any finite $\mu$.

We already know that, from the standpoint of the bulk theory, these kinks 
are confined monopoles \cite{SYnone,GSYmmodel}. The fact that tensions of all elementary strings
are the same ensures that these monopoles are free to move along the string,
since with their separation increasing, the energy of the configuration does not
change. This means they are in the
deconfinement phase.\footnote{We stress again that these monopoles are confined in the bulk theory being attached to strings.} The kinks are deconfined 
both in \ntwot and \ntwoo $CP(N-1)$ theories. In other words, individual kinks are present
in the physical spectrum. The monopoles although  attached to strings are 
free to move on the strings.

\begin{figure}
\epsfxsize=8cm
\centerline{\epsfbox{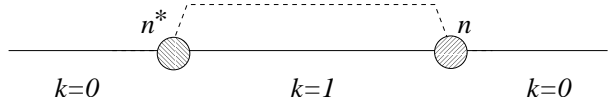}}
\caption{\small 
Linear confinement of the $n$-$n^*$ pair.
The solid straight line represents the ground state.
The dashed line shows
the vacuum energy density (normalizing ${\cal E}_0$ to zero).}
\label{fig:conf}
\end{figure}

The main distinction of the nonsupersymmetric $CP(N-1)$ model from its  
supersymmetric cousins is that the U(1) gauge field remains massless in the
absence of SUSY \cite{W79}.  The reason is that the nonsupersymmetric version does not have fermions 
$\xi^l$, while in the supersymmetric versions these fermions 
provide the photon with a mass  via the 
chiral anomaly. The presence of massless photon ensures long range forces in the 
nonsupersymmetric $CP(N-1)$ model. The Coulomb potential is linear in two dimensions
leading to the Coulomb/confinement phase \cite{W79}. Electric charges are confined.
The lightest electric charges are the  $n^l$ kinks. Confinement of kinks means that they are not
present in the physical spectrum of the theory in isolation. They form bound states,
kink-antikink ``mesons.'' The picture of
confinement of $n$'s is shown in Fig.~\ref{fig:conf}.

The validity of the above consideration rests on large $N$.
If $N$ is not large the solution \cite{W79}
ceases to be applicable. It remains  valid in the qualitative sense,
however. Indeed, at $N=2$ the model was solved exactly \cite{ZZ,Zam} (see also
\cite{Coleman}).
Zamolodchikovs found that the spectrum of the
O(3) model consists of a triplet of degenerate states
(with mass $\sim \Lambda$).
At $N=2$ the action (\ref{cpg}) is built of doublets.
In this sense one can
say that Zamolodchikovs' solution exhibits confinement of
doublets. This is in qualitative accord with the large-$N$ solution
\cite{W79}.

Inside the $\bar n\,n$ mesons, we have  a constant electric field,
see Fig.~\ref{fig:conf}. Therefore the spatial interval between $\bar n$
and $n$ has a higher energy density than the domains outside the meson.

Let us reiterate the above picture in somewhat 
different terms \cite{MMY,GSY05}. In the nonsupersymmetric model $N$ degenerate vacua
present in supersymmetric versions of the theory are split.
At large $N$,
along with the unique ground state,
the model has $\sim N$ quasistable local minima, quasivacua,
which become absolutely stable at $N=\infty$. The relative
splittings between the values of the energy density in the adjacent
minima
is of the order of $1/N$, while
the probability of the false vacuum decay is proportional to
$N^{-1}\exp (-N)$ \cite{nvacym,nvacymp}. 

The $n$
quanta (kinks) interpolate between the adjacent vacua.
They are confined monopoles of the bulk theory. Since the excited string tensions 
are larger than the tension of the lightest one, these monopoles,
besides four-dimensional confinement, are confined also in the two-dimensional sense:
a monopole is necessarily attached to an antimonopole on the string to form a meson-like 
configuration \cite{MMY,GSY05}. Otherwise, the energy of the configuration
will be infinitely higher (in a linear manner).

\section{ Conformal fixed point}
\label{conformal}
\setcounter{equation}{0}

In this section we discuss what happens if we send $\mu$ to infinity in the bulk theory.
This issue was addressed in \cite{Tongd} where it was 
 argued that in the 
$\mu\to\infty$ limit the $\sigma$-field
VEV vanishes, the U(1) gauge field is massless  and the 
theory is in the Coulomb/confinement phase (much in the same way as in
nonsupersymmetric $CP(N-1)$ models \cite{W79,GSYphtr}). 

In the $\mu\to\infty$ limit the adjoint fields decouple and  the bulk theory
flows to \none SQCD. It is well known that this theory has a Higgs branch
see, for example, \cite{IS}.
As was explained in \cite{SYnone}, the presence of the Higgs branch in the 
$\mu\to \infty$ limit is quite an unpleasant feature of the bulk theory.
The presence of massless states 
associated with the Higgs branch obscures physics of the non-Abelian strings. 
In particular, the strings swell and become infinitely thick.
This means that higher derivative corrections in the effective theory on the string become
important.
In \cite{SYnone} the maximal critical value of the parameter $\mu$ was estimated
beyond which one can no longer trust the effective two-derivative theory on 
the string worldsheet,
\beq
g^2_2\mu \ll \frac{m_W^{3}}{(\Lambda_{\, \rm bulk}^{\,{\mathcal N}=1})^2}\,,
\label{limit}
\eeq
where $\Lambda_{\rm bulk}^{{\mathcal N}=1}$ is the scale of  \none SQCD. We assume
week coupling in the bulk theory, i.e. $m_W\gg\Lambda_{\rm bulk}^{{\mathcal N}=1}$. 

Thus, we cannot go to the limit $\mu\to\infty$, to begin with. Higher derivative
corrections to the  worldsheet theory (\ref{cpg02}) blow up.
We   still have a large window in the  values of the  $\mu$ parameter,
with $\mu$ staying below the upper bound (\ref{limit}), but, on the other hand, 
large enough to ensure the decoupling of the adjoint fields, namely,
\beq
m_W\ll g^2_2\mu \ll m_W\, \frac{m_W^{2}}{\Lambda^2_{{\mathcal N}=1}}\,.
\label{window}
\eeq
Inside this window the deformation parameter $u$ is finite (see (\ref{umularge})). 
Our results show that  the $\sigma$-field VEV 
does not vanish and  we  have $N$ strictly degenerate vacua. 
Moreover,  the 
U(1) gauge field always has a small mass implying that the  kinks are in the deconfinement phase.
As we explained above the  mass generation for the photon field is in one-to-one
correspondence with the existence of $N$ distinct {\em degenerate}
vacua.

The situation with the decoupling of the adjoint fields we encounter here seems
counterintuitive, at least
at first sight. Indeed, on physical grounds we can say that
once $g_2^2\mu$ becomes larger than $m_W$ by a factor of, say,  5 or so
the adjoint fields are already decoupled, and the subsequent evolution of
their mass from $5m_W$ to infinity
should have no impact in the bulk as well as on the string worldsheet.
However, Eq.~(\ref{omegasmu}) shows that this is not the case.
The logarithmic growth at large $\mu$
seems to be a typical massless particle effect.
If the theory had no Higgs branch in the limit of the large bulk deformation parameter,
one can expect the worldsheet deformation parameter to be frozen at 
a finite value. We {\em conjecture} that that's what happens in the $M$ model \cite{GSYmmodel}.
In this model the Higgs branch does not develop.

Now, let us abstract ourselves from the fact
that the theory (\ref{cpg02}) is  a low-energy effective
model on the worldsheet of the non-Abelian string. Let us consider this model
{\em per se}, with no reference to the underlying
four-dimensional theory. Then, of course, the parameter $u$ can be viewed
as arbitrary. One can address a
subtle question: what happens in the limit $u\to\infty$? In this limit the  $\sigma$ field VEV
tends to zero (see Eq.~(\ref{sol})) and $N$ degenerate vacua coalesce. Moreover, the
U(1) gauge field, $\sigma$ and the fermionic kinks $\xi$ become massless (in addition to the
$\lambda_R$ field which, being  Goldstino in this limit, is necessarily massless). The low-energy theory seemingly becomes
conformal. It is plausible to interpret this conformal fixed point as a phase
transition point from the kink deconfinement phase to the Coulomb/confining phase.

A similar phenomenon occurs in two-dimensional conformal ${\mathcal N}=(4,4)$ supersymmetric
gauge theory  \cite{W97}. In this theory the same tube metric $|d\sigma|^2/|\sigma|^2$  appears 
(as in 
(\ref{41}), (\ref{esigma})) and the point
$\sigma=0$ is interpreted as a transition point between two distinct phases.

\section{Conclusions}
\label{conclu}

In this paper we discussed dynamics of the heterotic \ntwoo
$CP(N-1)$ model. Besides all fields of the conventional \ntwot
$CP(N-1)$ model the heterotic one contains a single extra right-handed fermion
$\zeta_R$. Interaction of the latter with other fields is characterized by a single 
dimensionless parameter (\ref{u}) which grows logarithmically with $\mu$,
 see Eq.~(\ref{umularge}). Using the large-$N$ expansion, we solved the 
heterotic model
in the leading order of this expansion.
The proof of the spontaneous supersymmetry breaking which for small
deformation parameters was  given in \cite{SY02}
is extended to arbitrary values of the
deformation parameter. We find the vacuum energy density for $N$ degenerate vacua present in the model. Lifting the vacuum energy from zero makes the \ntwoo model
akin to nonsupersymmetric $CP(N-1)$ model. 

The $Z_{2N}$ symmetry is broken down to $Z_2$
much in the same way as in the \ntwot model.
The vacua are labeled by the nonvanishing expectation values $\langle \sigma\rangle$,
see Eq.~(\ref{sol}), in the allowed window (\ref{window}) of the
values of the deformation parameter $u$.
The presence of $N$ distinct degenerate vacua
guarantees the theory to be in the deconfining phase. Correspondingly, the
mass of the two-dimensional photon is nonvanishing. This makes the \ntwoo model
akin to the \ntwot model.

We found the mass spectra at small and large values of
the deformation parameter. The small-$\mu$ case is rather
selfevident. At large $\mu$ we encounter a rather intriguing
situation: the only field whose mass is $\sim \Lambda$ is the $n$
field. Others are either much lighter or much heavier.

\section*{Acknowledgments}

We are grateful to S. Bolognesi for stimulating discussions.
Special thanks go to to A. Monin and P. Koroteev who pointed out to us (see \cite{monin})
that the wave function renormalization for Im$\sigma$ is different from that for
Re$\sigma$, 
the fact which we initially  overlooked. 
The impact of this is that at large $u$ the mass of the real part of $\sigma$ in (\ref{smass02}) is a factor of $\sqrt 3$ larger than that given in Version 1.

This work  is supported in part by DOE grant DE-FG02-94ER408. 
The work of A.Y. was  supported 
by  FTPI, University of Minnesota, 
by RFBR Grant No. 06-02-16364a 
and by Russian State Grant for 
Scientific Schools RSGSS-11242003.2.

\end{document}